# The Smart Data Extractor, a clinician friendly solution to accelerate and improve the data collection during clinical trials

Sophie QUENNELLE[a,c,d1], Maxime DOUILLET[b], Lisa FRIEDLANDER[c,d], Olivia BOYER[c,d], Antoine NEURAZ[a,c,d], Anita BURGUN[a,c,d], Nicolas GARCELON[a,b,d].

[a] *HeKA Team, Inria Inserm UMR_S1138, PariSantéCampus, Paris, France*
[b] *Data Science Platform, Imagine Institute, Paris, France*
[c] *Hôpital Universitaire Necker-Enfants malades, APHP, Paris, France*
[d] *Université de Paris Cité, Paris, France*

**Abstract.** In medical research, the traditional way to collect data, *i.e.* browsing patient files, has been proven to induce bias, errors, human labor and costs. We propose a semi-automated system able to extract every type of data, including notes. The Smart Data Extractor pre-populates clinic research forms by following rules. We performed a cross-testing experiment to compare semi-automated to manual data collection. 20 target items had to be collected for 79 patients. The average time to complete one form was 6'81'' for manual data collection and 3'22'' with the Smart Data Extractor. There were also more mistakes during manual data collection (163 for the whole cohort) than with the Smart Data Extractor (46 for the whole cohort). We present an easy to use, understandable and agile solution to fill out clinical research forms. It reduces human effort and provides higher quality data, avoiding data re-entry and fatigue induced errors.

**Keywords.** Electronic Health Records, Clinical Research Forms, Clinical Data Reuse, Observational Study

## 1. Introduction

Most of the patients' information required for clinical trials and registries are available in patients' electronic health records (EHRs).[1] The most common manner to fill Case Report Forms (CRF) is still to browse patients' documents searching for the information required by the study protocol. This process induces delays, human efforts, costs, and risks of transcription errors. Recent efforts have been dedicated to reuse EHR data to identify patients eligible for trials to optimize clinical trial protocols and to transcribe the variables of interest from EHRs to CRFs automatically.[2, 3] However, several pitfalls remain since EHR data are heterogeneous, completeness of structured data elements is low and most of the clinical information is locked into medical notes and needs to be transformed in a structured format before secondary use.[4] Our objective was to develop a pipeline able to speed up the collection of data required by a CRF, from all document sources, and to support user-friendly data quality assessment. We evaluated this tool through a retrospective study.

---

[1] Corresponding Author : Sophie Quennelle, sophie.quennelle@protonmail.com

## 2. Method

*2.1. Material*

Necker-Enfants Malades Hospital is an AP-HP university children's hospital in Paris with a data analytics and warehousing solution, called Dr. Warehouse (DRWH) [5]. It integrates multiple data sources ranging from structured data to free text clinical narratives and applies natural language processing methods to medical text to detect negation, family history, and to extract phenotypic information based on the Unified Medical Language System Metathesaurus®.

*2.2. Methods*

The Smart Data Extractor (SDE) has been developed on top of DRWH to help researchers with patient information retrieval and CRF completion. The SDE is adapted as follows to populate a given CRF (Figure 1): (1) The user provides a formal representation of each item of the research protocol (name, type, list of accepted values) and associates specific extraction rules to each item. The items and queries can be imported or created by an expert. If the data of interest is stored in a structured format, the rule includes the corresponding thesaurus codes (e.g. LOINC, ICD10, or local nomenclature).

If the data is to be searched for in the clinical notes, the system can simply find the documents containing the items based on the labels of the variables or items in a list and check the corresponding box. The user can also specify the regular expressions (REGEX) that are used to retrieve information from clinical notes. (2) The software works as an automatic search engine mining the patients reports to look for the sentences containing the items and values.

The SDE has been designed to be used in two ways: semi or fully automated. In the fully automated mode, the SDE extracts automatically the completed CRF for each patient of the study cohort. This approach is appropriate when the data of the questionnaire are unambiguous. In the semi-automated mode, the user needs to validate manually each item of the CRF and can modify the answers. This approach is suitable for ambiguous variables that require human validation. We designed a proper Human Machine Interface for this validation. For each item, the SDE screens the patient EHR and suggests variables matching the query. If an answer is found in a negative syntagm, the interface displays this information to the user. As the variables are presented in their context (sentences) the expert is able to select the most appropriate occurrences. The link between the data extracted and the health reports is maintained, to ensure data assessment at any time.

At any time, experts can edit and improve the list of items and queries of the SDE. When the CRF is completed for the cohort, it can be exported into a format suitable for statistical analysis. Authorized users can export the data and/or transfer it to a Research Electronic Data Capture (REDCap) database.[6]

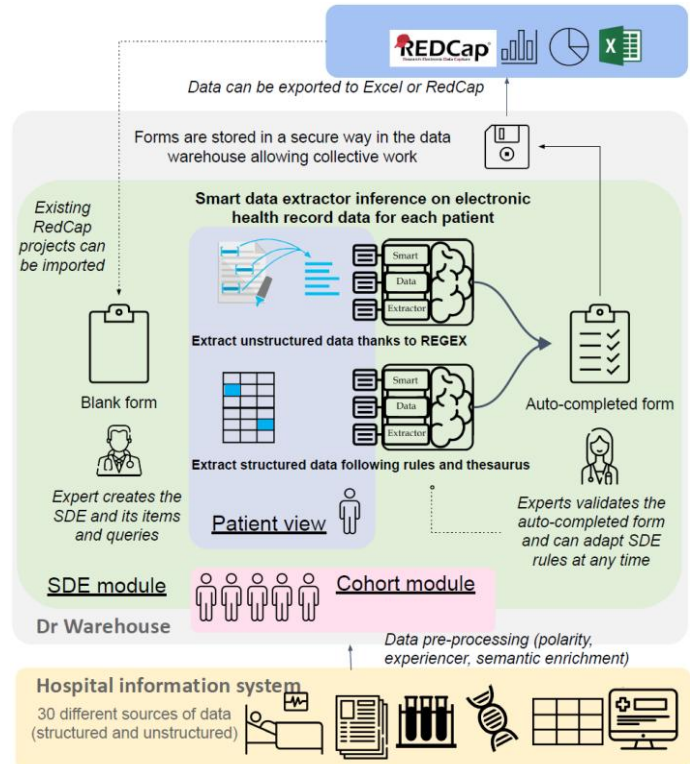

**Figure 1.** Graphical presentation of the Smart Data Extractor.

*2.3. Evaluation*

The evaluation protocol consisted in comparing manual versus SDE-assisted data collection of those 20 variables required to calculate a risk score related to the cardiac catheterization procedure. A medical doctor (MED) designed a SDE dedicated to this task after two days of training on REGEX and DRWH Thesaurus. Predefined functions were used for demographic data. Biological and hemodynamics were extracted based on the corresponding thesaurus code. A list of REGEX was defined for unstructured items. When appropriate, time intervals were defined in order to limit the data collection in a specific period around the procedure. We randomly divided 79 patients into two subgroups. Two researchers, a clinical research assistant with expertise in DRWH (TEC) and a cardiologist (MED) conducted the data collection. They filled in the forms in a cross-testing experiment: for the first subgroup, the research assistant completed the CRF manually, i.e., he read every patient's files on the database and filled in the eCRF without any computer assistance while the cardiologist used the SDE. For the second subgroup TEC was assisted by the SDE while MED completed the eCRF manually. The two cohorts were then reconciliated to identify discrepancies, compare the share of missing data and data collection times.

## 3. Results

Group 1 was composed of 39 patients, with a mean number of documents of 111 and a mean length of follow-up of 2.8 years. Group 2 was composed of 40 patients, with a mean number of documents of 177 and a mean length of follow-up of 3.2 years. The mean data collection time per patient was lower when the users were assisted by the SDE (3'27'' for TEC + SDE on group 2 and 3'17'' for MED + SDE on group 1 versus 6'23'' for TEC alone on group 1 and 7'38'' for MED alone on group 2).

Discrepancies between TEC and MED were comparable for the two groups (103 for group 1 and 106 for group 2). Regardless of the user, there were more mistakes when the form was filled in manually (in total, 163 errors in the manually filled in forms versus 46 in the semi-automated filled in forms). Three types of errors were made: (1) missing values, i.e. data that were not retrieved in patient EHR by one of the researchers, (2) breach of data collection protocol, (3) misunderstanding of an ambiguous type of procedure or of the patient's medical history in text.

We paid attention to the software's display and ergonomics. The patient file and the CRH are displayed side by side allowing the user to read the context of each extracted concept. TEC appreciates the fact that the SDE points out the relevant part of the patient file for each clinical question. MED highlighted the fact that the SDE was able to detect family history and negation which avoids the burden generated through false positives keywords detection.

## 4. Discussion

We presented and evaluated the SDE, a generic inference engine that automatically populates CRF based on EHRs. Globally, the time needed to fill out the CRF was divided by two and the number of errors by three. The results were similar for the two users (MED or TEC).

Medical data extraction solutions should permit text mining as clinical narratives are the primary source of information about patient history, treatment, and disease course. To avoid the burden of false positive keywords, it is critical to detect negated clinical signs and family medical history. Both automatic and semi-automatic data extraction can lead to errors [7] but the semi-automated method reduces the eventuality that the researchers make two different interpretations of the exact same event, the final human check ensures that the algorithm is working properly and requires fewer human and computer resources to develop [8]. With the SDE the link between each extracted information and its document source and context is maintained, which ensures the possibility of data verification and traceability at any time. A clinical extraction tool must be easily adaptable to the needs of any research protocol while remaining simple enough to be handled by clinicians. Shalhout and al. proposed a pipeline to capture structured clinical data to a REDCap based registry.[9] Miller and al. provide a powerful application to abstract clinico-genomic data from EHR but this extraction solutions are limited to a predefined sets of data.[10] We sought to design the SDE in an intuitive and user centered way so that non informatic trained researchers (nurses, medical students, physicians…) could configure the data collection forms according to their specific needs.

Our evaluation protocol had some limitations: first, it is not possible to discriminate with certainty missing data from data that existed in the EHR but was not identified by either of the two users. Secondly, the time allocated to implement the tool has not been measured. Thirdly, the rules, conceived by a non-specialist, were probably not optimal. This may have had an impact on the efficiency of our SDE, but it also reflects the real-life use of the tool. Finally, in our specific use case 5 variables only had to be searched in the text which explains the relatively short time needed to complete each patient form. A more tedious CRF would probably have shown a more significant gain with the SDE. To favor the widespread use of the tool, efforts should be made on interoperability. We aim to make the SDE universal and adaptable to any type of hospital database. This could facilitate secure data export and exchange and so multi-center clinical registries.

## 5. Conclusion

The SDE is an easy-to-use semi-automated data collection system able to extract all variables of interest whatever their format (structured and unstructured) from the patient EHR to fill in CRF. The SDE automatically extracts patients' structured data and assists researchers in text mining for the semi-automatic extraction of data reported by caregivers in clinical notes. We are convinced that the SDE can promote multi-centered trials, reduce costs and clinical research cycle time. The SDE is a semi-automated framework, it requires human effort and validation and does not guarantee zero missing data and error rates but unlike a complex NLP model, it does not necessitate a step of training and can adapt to any clinical subject in a short time.

## References

[1]   A. El Fadly, B. Rance, N. Lucas, C. Mead, G. Chatellier, P.-Y. Lastic, M.-C. Jaulent, and C. Daniel, Integrating clinical research with the Healthcare Enterprise: From the RE-USE project to the EHR4CR platform. Journal of Biomedical Informatics. 2011 Jul. S94–S102, doi:10.1016/j.jbi.2011.07.007.
[2]   EHR4CR | Electronic Health Records Systems for Clinical Research, IMI Innovative Medicines Initiative.http://www.imi.europa.eu/projects-results/project-factsheets/ehr4cr
[3]   G.A. Brat, G.M. Weber, and al, The Consortium for Clinical Characterization of COVID-19 by EHR (4CE), T. Cai, and I.S. Kohane. International Electronic Health Record-Derived COVID-19 Clinical Course Profiles: The 4CE Consortium, Infectious Diseases (except HIV/AIDS). 2020 Apr. doi:10.1101/2020.04.13.20059691.
[4]   A. Vass, I. Reinecke, M. Boeker, H.-U. Prokosch, and C. Gulden. Availability of Structured Data Elements in Electronic Health Records for Supporting Patient Recruitment in Clinical Trials, Stud Health Technol Inform. 2022 Jun. 130–134. doi:10.3233/SHTI220046.
[5]   Garcelon N, Neuraz A, Salomon R, Faour H, Benoit V, Delapalme A, Munnich A, Burgun A, Rance B. A clinician friendly data warehouse oriented toward narrative reports: Dr. Warehouse. J Biomed Inform. 2018 Apr. doi: 10.1016/j.jbi.2018.02.019.
[6]   P.A. Harris, R. Taylor, R. Thielke, J. Payne, N. Gonzalez, and J.G. Conde. Research Electronic Data Capture (REDCap) - A metadata-driven methodology and workflow process for providing translational research informatics support/ J Biomed Inform. 2009 Apr. 377–381. doi:10.1016/j.jbi.2008.08.010.
[7]   M.D. Byrne, T.R. Jordan, and T. Welle. Comparison of Manual versus Automated Data Collection Method for an Evidence-Based Nursing Practice Study. Appl Clin Inform. 2013 Jan. 61–74. doi:10.4338/ACI-2012-09-RA-0037.
[8]   REDCap on FHIR: Clinical Data Interoperability Services, Journal of Biomedical Informatics. 2021 May. 103871. doi:10.1016/j.jbi.2021.103871.
[9]   S.Z. Shalhout, F. Saqlain, K. Wright, O. Akinyemi, and D.M. Miller. Generalizable EHR-R-REDCap pipeline for a national multi-institutional rare tumor patient registry. JAMIA Open. 2022 Sep. ooab118. doi:10.1093/jamiaopen/ooab118.
[10]  D.M. Miller, and S.Z. Shalhout. GENETEX-a GENomics Report TEXt mining R package and Shiny application designed to capture real-world clinico-genomic data. JAMIA Open. 2021 Jul. ooab082. doi:10.1093/jamiaopen/ooab082.